\documentclass[prl,superscriptaddress,showpacs,longbibliography,reprint]{revtex4-2}

\usepackage{hyperref}

\usepackage{color}
\usepackage[usenames,dvipsnames]{xcolor}
\usepackage{amsmath,amsthm,amssymb}
\usepackage{graphicx}
\usepackage{epsfig}
\usepackage{dcolumn}
\usepackage{bm}
\usepackage{mathrsfs}
\usepackage{multirow}
\usepackage[all]{xy}
\usepackage{pbox}
\usepackage{verbatim}
\usepackage{braket}
\usepackage{dsfont}

\usepackage{graphicx}
\graphicspath{{./Figures/}}

\usepackage{color,soul}

\begin{document}

\title{Entangled multiplets and unusual spreading of quantum correlations in a continuously monitored tight-binding chain}
\author{Federico Carollo}
\affiliation{Institut f\"{u}r Theoretische Physik, Universit\"{a}t T\"{u}bingen, Auf der Morgenstelle 14, 72076 T\"{u}bingen, Germany}
\author{Vincenzo Alba}
\affiliation{Dipartimento di Fisica, Universit\`a di Pisa, and INFN Sezione di Pisa, Largo Bruno Pontecorvo 3, Pisa, Italy}

\begin{abstract}
We analyze the dynamics of entanglement in a paradigmatic noninteracting system subject to continuous 
monitoring of the local excitation densities. 
Recently, it was conjectured that the evolution of 
quantum correlations in such system is described by a semi-classical theory, based on entangled pairs of ballistically propagating quasiparticles and inspired by the hydrodynamic approach to unitary (integrable) quantum systems. 
Here, however, we show that this conjecture does not fully capture the complex behavior of quantum correlations emerging from the interplay between coherent dynamics and continuous monitoring. 
We unveil the existence of multipartite quantum correlations which are inconsistent with an entangled-pair structure and which, within a quasiparticle picture, would  require the presence of larger multiplets. We also observe that quantum information is highly delocalized, as it is shared in a collective {\it nonredundant} way among adjacent regions of the many-body system. 
Our results shed new light onto the behavior of  correlations in quantum stochastic 
dynamics and further show that these may be enhanced by a (weak) continuous monitoring process.
\end{abstract}

\maketitle

The evolution of quantum correlations in stochastic systems is attracting much attention nowadays \cite{nahum2017,nahum2018,zhou2019,cao2019,znidaric2020,piroli2020,nahum2021,alberton2021,lavasani2021,ippoliti2021,coppola2021}. 
On one hand, this dynamics is relevant for understanding the extent to which quantum effects may be exploited in current devices  \cite{peruzzo2014,preskill2018,kandala2019,havlivcek2019,guerreschi2019,arute2019,zhong2020,xia2021,bharti2022}. 
On the other, this renewed interest has been triggered by the discovery of entanglement 
phase transitions \cite{li2018,skinner2019,chan2019,jian2020,gullans2020,bao2020,lang2020,rossini2020,tang2020,lunt2020,zabalo2020,lunt2021,turkeshi2021,buchhold2021,sang2021,lu2021,jian2021,agrawal2021,block2022,minato2022,muller2022,minoguchi2022,zabalo2022}, stemming from the competition between coherent dynamics and random measurements. Furthermore, 
quantum stochastic processes hold the promise to bridge recent progress in the description of nonequilibrium unitary quantum systems and open challenges in characterizing 
open quantum dynamics \cite{lange2018,rossini2021,rossini2021b,maity2020,bouchoule2020,alba2021,alba2021c,carollo2022,alba2021b,starchl2022}, 
also beyond average-state properties \cite{carollo2019,carollo2020,bernard2021,turkeshi2022}. In this regard, 
demonstrating the applicability of powerful theories, so-called quasiparticle pictures \cite{calabrese2005,fagotti2008,alba2017,calabrese2018}, 
to entanglement spreading in stochastic many-body processes would represent a major breakthrough. 

\begin{figure}[t]
\centering
\includegraphics[width=\linewidth]{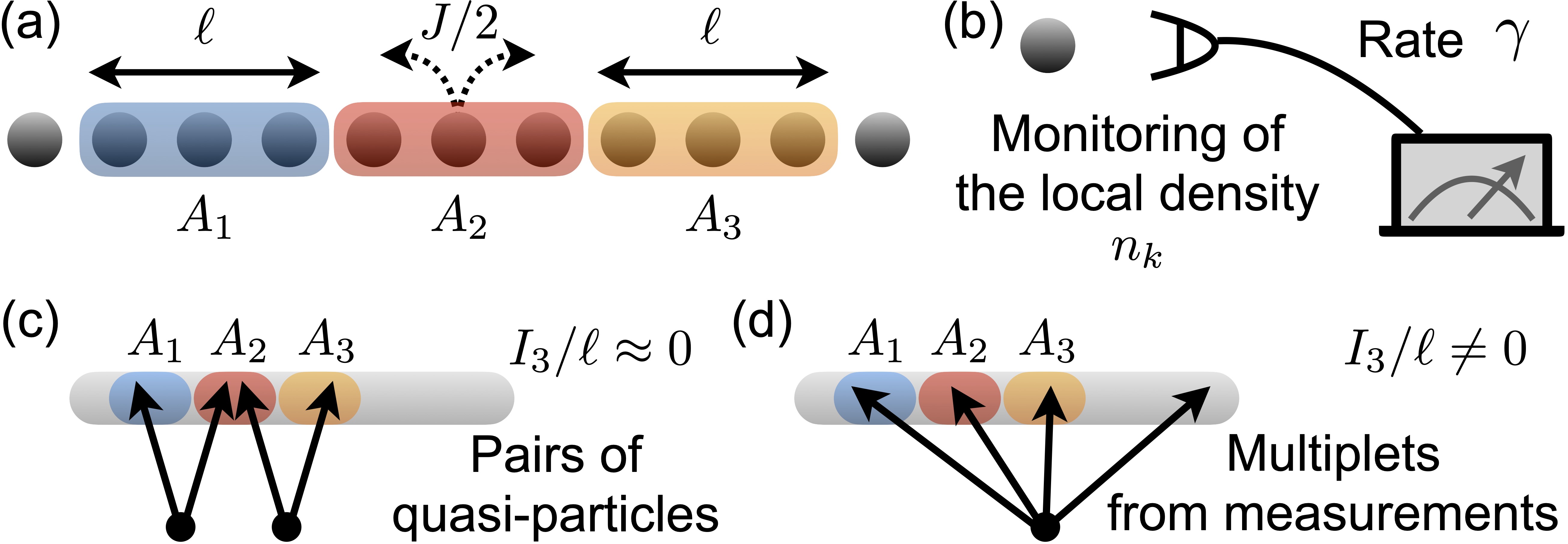}
\caption{{\bf Noninteracting system subject to continuous monitoring.} a) Fermionic tight-binding chain 
with coherent hopping rate $J/2$. We consider several partitioning of this many-body system, $A\cup\bar{A}$, into a system of interest $A$ and its complement, $\bar{A}$. In the sketch, we illustrate a system $A$ made of three adjacent subsystems, $A=A_1\cup A_2\cup A_3$, of equal length $\ell$. b) Each site of the chain is subject to the continuous measurement of its local density $n_m$, at rate $\gamma$. c) The tripartite mutual information $I_3$ quantifies the degree of extensivity of the mutual information and is also a fourpartite entanglement measure for pure states \cite{pavan2016}. Pairs of quasiparticles cannot entangle more than two intervals at a same time, implying $I_3/\ell\to0$. d) Multiplets with at least four quasiparticles can lead to a finite tripartite mutual information $I_3/\ell$.
 }
\label{Fig1}
\end{figure}

This possibility has been explored in a paradigmatic many-body quantum system \cite{lieb1961}, subject to continuous monitoring \cite{wiseman2009} [see sketch in Fig.~\ref{Fig1}(a-b)]. It has been proposed that, as it happens for unitary (integrable) dynamics \cite{calabrese2005,fagotti2008,alba2017,calabrese2018}, the spreading of correlations in the system
is solely attributable to entangled pairs of ballistically propagating quasiparticles \cite{cao2019}. 
The effect of continuous monitoring was conjectured to be that of making these excitations unstable and of generating new entangled pairs, in place of the collapsed ones \cite{cao2019}. 
This {\it collapsed quasiparticle ansatz} \cite{cao2019} has been benchmarked against exact numerics for the von Neumann entanglement entropy, showing remarkable results \cite{cao2019,coppola2021}.

In this paper, however, we demonstrate that continuously monitored systems feature an unexpectedly complex dynamics of quantum correlations, whose fundamental aspects are not captured by the collapsed quasiparticle ansatz.   
We show indeed that such a theory is not quantitatively accurate in predicting several measures of bipartite entanglement, both between a subsystem and the remainder of the many-body system and between adjacent subsystems. In this latter setting, we observe that continuous monitoring can also be, quite surprisingly, beneficial for quantum correlations, since it can stabilize a stationary entanglement in cases in which the unitary dynamics would lead to unentangled subsystems.  

Most importantly, we identify a central reason why the dynamics of quantum correlations in the system cannot be captured by a picture based on quasiparticle pairs.  We compute the {\it tripartite mutual information} between three subsystems $A_1,A_2,A_3$ [see sketch in Fig.~\ref{Fig1}(a)] and show that it assumes nonzero,  in fact  negative, values. This signals the existence of multipartite (i.e., between more than two intervals)
quantum correlations, which are inconsistent  
with the mere presence of pairs of entangled quasiparticles [cf.~Fig.~\ref{Fig1}(c)]. As we discuss,
consistency of a quasiparticle picture with a nonzero tripartite mutual information requires the presence of entangled multiplets with at 
least four quasiparticles, as sketched in Fig.~\ref{Fig1}(d).  

A negative tripartite mutual information, as we find here, implies that  the information about $A_2$ contained in $A_1\cup A_3$ is more than the sum of the information contained in $A_1$ and $A_3$ separately \cite{iyoda2018}, showing that for the monitored system {\it the whole is more than the sum of its parts} \cite{seshadri2018}. It further indicates that the mutual information is monogamous  and, thus, likely to be dominated by quantum correlations \cite{hayden2013,rota2016,asadi2018,seshadri2018,ali-akbari2021}. Negative values of the tripartite mutual information have also been associated with the delocalization (broadly referred to as {\it scrambling}) of quantum information \cite{pavan2016,seshadri2018,iyoda2018,schnaack2019,nie2019,landsman2019,ali-akbari2021,blok2021,kudler-flam2022,zhu2022}. Our findings show that the interplay between monitoring and coherent dynamics leads to the continuous dispersal of quantum information into entangled multiplets of excitations, which in turn establish robust multipartite entanglement and determine an unusual, for a noninteracting system, dynamics of quantum correlations.  \\

\noindent {\bf Monitored noninteracting system.---} We consider a   fermionic chain subject to the continuous measurement of local 
observables \cite{cao2019,alberton2021}. The model Hamiltonian is 
\begin{equation}
H=\frac{J}{2}\sum_{m=1}^L \left(a^\dagger_m a_{m+1}+a_{m+1}^{\dagger }a_m\right)\, ,
\label{Ham}
\end{equation}
with $a_m,a_m^\dagger$ being fermionic annihilation and creation operators. 
This Hamiltonian describes coherent hopping of fermionic excitations between neighboring sites at rate $J/2$ [cf.~Fig.~\ref{Fig1}(a)], 
in a periodic lattice. 
The total number of fermionic excitations $N=\sum_{m=1}^L n_m$, with $n_m=a^\dagger_m a_m$, is conserved and we assume that the local 
fermionic densities $n_m$ are continuously measured, as sketched in Fig.~\ref{Fig1}(b). 
This monitoring induces non-linear and random effects in the 
system dynamics, which is governed by the stochastic Schr\"odinger equation  \cite{breuer02a,gardiner2004,wiseman2009,barchielli2009}
\begin{multline}
d\ket{\psi(t)}=-iHdt\ket{\psi(t)}\\
+\sum_{m=1}^L \!\left(\sqrt{\gamma} M_m(t) dW_{m}(t)-\frac{\gamma}{2} M_m^2(t) dt\right)\!\ket{\psi(t)}\, ,
\label{eq-sto}
\end{multline}
where $M_m(t)=n_m-\bra{\psi(t)}n_m\ket{\psi(t)}$. The terms $dW_{m}(t)$ are Wiener 
processes --- in Ito convention --- such that $\mathbb{E}[dW_{m}(t)]=0$ and 
$\mathbb{E}[dW_{m}(t)dW_{m'}(t)]=\delta_{mm'}dt$, with $\mathbb{E}$ denoting 
expectation over noise realizations. The rate $\gamma$ provides the strength of the monitoring process. 

We consider the initial state to be the N\'eel state 
$\ket{\psi(0)}=\prod_{ m\, {\rm odd } }a_m^\dagger \ket{0}$, 
where $\ket{0}$ is the fermionic vacuum. For each noise realization, Eq.~\eqref{eq-sto} 
encodes a quantum trajectory. Since the initial state is Gaussian and the generator is quadratic, the state in each trajectory is Gaussian at all times and can be efficiently simulated \cite{cao2019,alberton2021}.  In particular, entanglement-related 
quantities, such as the R\'enyi entropies $S^{(n)}_\ell(t)$ of a subsystem 
of length $\ell$, in quantum trajectories can be calculated from the fermionic two-point function 
$C_{hk}=\langle\psi(t)| a^\dagger_h a_k|\psi(t)\rangle$ \cite{peschel2009}. 
In what follows, we focus on the behavior of the entropies $\overline{S}^{(n)}_\ell(t):=\mathbb{E}[S^{(n)}_\ell(t)]$, and related quantities, averaged over quantum trajectories. \\

\noindent {\bf Collapsed quasiparticle ansatz.---} 
In the absence of continuous monitoring ($\gamma\equiv0$), the unitary dynamics of quantum information in the system is 
captured by a quasiparticle picture \cite{calabrese2005,fagotti2008,alba2017,calabrese2018}. 
The basic idea is that the initial state of the system acts as a source of pairs of entangled quasiparticles, labelled by 
their quasi-momentum $q$, which travel ballistically in opposite direction with 
velocity $|v_q|=|J\sin(q)|$. While travelling, quasiparticles  
spread correlations along the system. Specifically, the entanglement between a subsystem and its complement, at a given time, is proportional to 
the number of quasiparticle pairs they share at that time. 
For instance, the R\'enyi-$n$ entanglement entropy of a subsystem of length $\ell$, embedded in an infinite chain, is given by~\cite{fagotti2008}
\begin{equation}
	S^{(n),0}_\ell(t)=\int_{-\pi}^{\pi}\frac{dq}{2\pi} \, 
	s^{(n)}_q \Theta_q(t)\, .
\label{entropy-unitary} 
\end{equation}
This equation is valid in the scaling limit $t,\ell\to\infty$, with $t/\ell=\tau$ fixed, and provides the leading-order behavior in $\ell$. 
The superscript $0$ stands for unitary dynamics and 
\begin{equation}
	\Theta_q(t)=\mathrm{min}\{2|v_q|t,\ell\}\, .
\end{equation}
This function counts the number of pairs, formed by 
quasiparticles with quasimomenta $q$ and $-q$, 
shared by the subsystem and its complement at time $t$~\cite{calabrese2005}. The term 
$s_q^{ (n)}$ accounts for the entanglement between such quasiparticles and is given by the Yang-Yang entropy 
\begin{equation}
s^{(n)}_{q}=(1-n)^{-1}\log \left[\varrho_q^n+(1-\varrho_q)^{n}\right]\, ,
\label{eq:gge}
\end{equation}
quantifying the quasimomentum
contribution to the thermodynamic entropy of the generalized Gibbs ensemble describing local stationary properties of the system \cite{polkovnikov2011,caux2013,caux2016,calabrese2016,essler2016,vidmar2016}. 
In Eq.~\eqref{eq:gge}, $\varrho_q$ is the density of quasiparticles 
$\varrho_q=\bra{\psi(0)}\beta_q^\dagger \beta_q\ket{\psi(0)}$ and $\beta_q$ are  the eigenmodes of the Hamiltonian $H$. For the N\'eel state, $\varrho_q=1/2$, $\forall\, q$. 

\begin{figure*}[t]
\centering
\includegraphics[width=\linewidth]{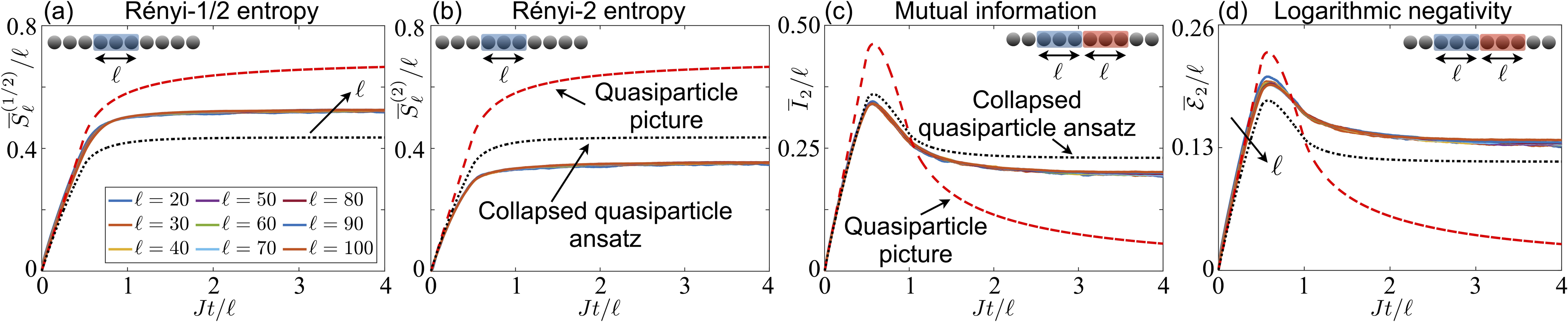}
\caption{{\bf Entanglement and quantum correlations in the monitored system. } a) Average R\'enyi-$1/2$ entropy for a subsystem of length $\ell$ --- see sketch --- quantifying entanglement between the subsystem and its complement (the remainder of the system). The dashed line is the unitary prediction from  Eq.~\eqref{entropy-unitary}, while the dotted line is the one from Eq.~\eqref{entropy-stochastic}. Solid lines are numerical results. We have taken $\Gamma/J=1$. (b) Same as in (a) but for the R\'enyi-$2$ entanglement entropy. The predictions coincide with the ones in (a), see main text.  (c) Mutual information $\overline{I}_2$ [cf.~Eq.~\eqref{mutual_information}] between two adjacent subsystems of length $\ell$, as shown in the sketch. Also in this case, the dashed line is the prediction for the unitary case $\gamma\equiv0$, while the dotted one is the prediction from the collapsed quasiparticle ansatz (see Ref.~\cite{SM}). (d) Logarithmic negativity $\overline{\mathcal{E}}_2$ quantifying entanglement between two adjacent subsystems. Both predictions coincide with half of those obtained for the mutual information \cite{SM}. For all panels, we considered $L=1000$. In (a-b), we averaged over $N_{\rm traj}=250$  trajectories, in (c)  $N_{\rm traj}=150$ while in (d) $N_{\rm traj}=100$.  }
\label{Fig2}
\end{figure*} 

To account for the presence of continuous monitoring, a modification to the above picture, also called 
collapsed quasiparticle ansatz, has been proposed \cite{cao2019}. 
The key assumptions are that quantum correlations are 
still exclusively spread by pairs of quasiparticles and that the measurement process solely determines their random collapse, at a rate proportional to $\gamma$.
When such an event occurs, the collapsed pair becomes irrelevant. However, in its place, a new entangled pair is produced, uniformly in quasimomentum. For (macroscopically) homogeneous initial states, such as the N\'eel state, it was assumed that the entanglement content of any pair, either generated in the initial state or during the dynamics, is a function of the average density \cite{cao2019}, which  is conserved by Eq.~\eqref{eq-sto}. 

From this stochastic picture, one can make predictions for the R\'enyi entropy of a subsystem averaged over trajectories. For homogeneous initial states, one has \cite{cao2019} 
\begin{equation}
\overline{S}_\ell^{(n)}(t)=e^{-\gamma t}\, 
	S^{(n),0}_\ell(t)+\gamma \int_0^{t} \!\!du \, e^{-\gamma u}\, S^{(n),0}_\ell(u)\, .
\label{entropy-stochastic}
\end{equation}
As for the unitary case, this equation is expected to provide the leading-order behavior in the scaling limit 
$t,\ell\to\infty$, with $\tau=t/\ell$ fixed. Since we have $t\propto \ell$, to make Eq.~\eqref{entropy-stochastic} well-defined in the $\ell\to\infty$ limit, it is natural to consider a small $\gamma$, obtained through the rescaling
$\gamma=\Gamma/\ell$, such that $\gamma t=\Gamma \tau$ remains 
fixed in the limit~\cite{lange2018,cao2019,bouchoule2020,alba2021,alba2021c,carollo2022} (see Supplemental Material \cite{SM}).
The first term in Eq.~\eqref{entropy-stochastic} accounts for correlations 
due to quasiparticle pairs formed in the initial state and survived up  to time $t$.  The second term instead accounts for 
pairs generated after the random collapses \cite{cao2019}. For $\gamma\equiv 0$, one recovers the unitary case $\overline{S}_\ell^{(n)}(t)=S^{(n),0}_\ell(t)$. Since $s_q^{(n)}=\log 2$ $\forall n$, Eqs.~\eqref{entropy-unitary}-\eqref{entropy-stochastic} give the same quantitative prediction for all R\'enyi entropies. \\ 

\begin{figure*}[t]
\centering
\includegraphics[width=\linewidth]{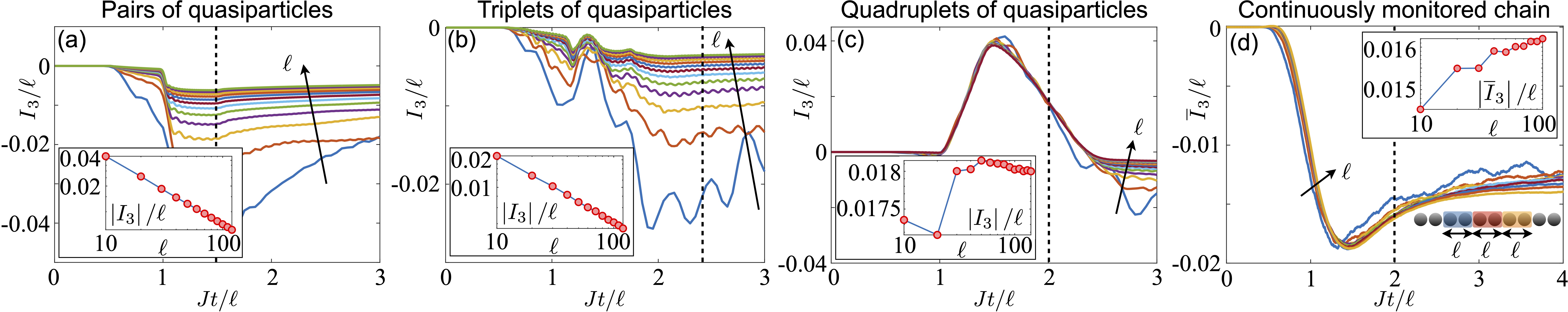}
\caption{{\bf Tripartite mutual information.} a) Dynamics of the tripartite mutual information starting from the N\'eel state, for $\gamma\equiv0$. In the scaling limit $\ell\to\infty$, this quantity is subextensive in $\ell$. The inset (in log-log scale) shows convergence to zero of $|{I}_3|/\ell$, for $Jt/\ell=1.5$. We considered $\ell=10,20,\dots 120$ and $L$ up to $L=1200$. b) Tripartite mutual information starting from a state with one fermionic excitation every three sites, for $\gamma\equiv0$. A subextensive behavior with $\ell$ of this quantity is apparent. The inset (in log-log scale) shows $|{I}_3|/\ell$ as a function of $\ell$, for $Jt/\ell=2.4$. We considered $\ell=10,20,\dots 120$ and $L$ up to $L=1200$.  c) Tripartite mutual information starting from a state with one fermionic excitation every four sites, for $\gamma\equiv0$. This quantity is extensive in $\ell$ and remains finite in the scaling limit, as also highlighted in the inset (in log-log scale) for $Jt/\ell=2$. We considered $\ell=10,20,\dots 140$ and $L$ up to $L=1500$. d) Average tripartite mutual information, $\overline{I}_3$, for the continuosly monitored system sketched in Fig.~\ref{Fig1}(a-b), with $\Gamma/J=1$. As shown in the main panel, as well as in the inset (in log-log scale) for $Jt/\ell=2$, the tripartite mutual information remains finite (negative) in the scaling limit $\ell\to\infty$. We considered $\ell=10,20,\dots 100$ and $L$ up to $L=1000$. The value of $\overline{I}_3$ is obtained by averaging over $N_{\rm traj}=250$ quantum trajectories.}
\label{Fig3}
\end{figure*} 

\noindent {\bf Entanglement and quantum correlations.---} 
We first analyze entanglement between a subsystem of length $\ell$ 
and its complement (the remainder of the many-body system), as sketched in Fig.~\ref{Fig2}(a). 
We consider the R\'enyi-$1/2$ 
entanglement entropy of the subsystem, which for each quantum trajectory is {\it exactly} equal to the logarithmic negativity \cite{peres1996,zyczkowski1998,eisert1999,vidal2002,shapourian2017,shapourian2019,shapourian2019b,gruber2020}, since the system state is pure \cite{calabrese2013entanglement}. As shown in Fig.~\ref{Fig2}(a), 
numerical results for  $\overline{S}_\ell^{(1/2)}$ do not agree with 
the prediction from Eq.~\eqref{entropy-stochastic} (dotted line). This also happens 
for the R\'enyi entropy with $n=2$ [see Fig.~\ref{Fig2}(b)]. As reported in Ref.~\cite{SM}, we even observe discrepancies between numerical results and the prediction for the von Neumann entropy analyzed in Ref.~\cite{cao2019}, when systematically considering the scaling limit. In Fig.~\ref{Fig2} (a-b), we also show the theory prediction 
for $\gamma\equiv 0$ (dashed line) given by Eq.~\eqref{entropy-unitary}. In the scaling limit $\overline{S}_\ell^{(n)}$ and ${S}_\ell^{ (n),0}$ are of the same order, even if the monitoring process suppresses here quantum correlations. 

We now consider bipartite correlations between two subsystems embedded in the chain. We start investigating the von Neumann mutual information, defined as
\begin{equation}
I_2[X,Y]:=S^{(n\to 1)}[X]+S^{(n\to 1)}[Y]-S^{(n\to 1)}[X\cup Y]\, ,
\label{mutual_information}
\end{equation} 
where $S^{(n\to 1)}[X]$ indicates the von Neumann entropy of the system $X$. In particular, we take two adjacent subsystems, $A_1$ and $A_2$, of length $\ell$ [see sketch in Fig.~\ref{Fig2}(c)]. Within the collapsed quasiparticle ansatz, the prediction for the average mutual information --- which we derived in Ref.~\cite{SM} ---  is given by an equation similar to Eq.~\eqref{entropy-stochastic}, with unitary term given by Eq.~\eqref{entropy-unitary} with $\Theta_q(t)=2|v_q| t+2\max\{|v_q| t,\ell\}-2\max\{2|v_q|t,\ell\}$ \cite{SM}. This function $\Theta_q(t)$ now counts the number of quasiparticle pairs shared 
by the intervals $A_1$ and $A_2$ \cite{coser2014}. 

As shown in Fig.~\ref{Fig2}(c), $\overline{I}_2(t)$ exhibits clear scaling behavior in the limit. Still, as for the entanglement entropies, the theoretical prediction fails to capture quantitatively the mutual information. For instance, our results show that correlations between $A_1$ and $A_2$ do not decay to zero in the limit $t/\ell\to\infty$, in contrast with the unitary case (dashed line), but reach a plateau value. While this feature is qualitatively captured by the ansatz \cite{SM}, the exact stationary value is substantially different. The existence of this plateau demonstrates that the continuous monitoring enhances bipartite correlations. This is due to the fact that the monitoring  generates, continuously in time, entangled excitations throughout the system, and their spreading sustains finite stationary correlations between the two subsystems. Since the quantum state of $A_1\cup A_2$ is mixed, these correlations are in principle both of quantum and of classical nature.  However, we can also calculate the logarithmic negativity $\overline{\mathcal{E}}_2(t)$ \cite{peres1996,zyczkowski1998,eisert1999,vidal2002,shapourian2017,shapourian2019,shapourian2019b,gruber2020}, a proper measure of entanglement, which shows that $A_1$ and $A_2$ are not solely classically correlated but feature a stationary entanglement, as shown in Fig.~\ref{Fig2}(d). The prediction from the collapsed quasiparticle ansatz for the logarithmic negativity is given by $\overline{\mathcal{E}}_2(t)=\overline{I}_2(t)/2$ \cite{SM}. This is due to the fact that for the unitary system the logarithmic negativity is equivalent to the R\'enyi-$1/2$ mutual information in the scaling limit \cite{alba2019b}, and that for our case the latter is equal to $I_2(t)$. The above prediction fails to capture the behavior of $\overline{\mathcal{E}}_2(t)$, as shown in Fig.~\ref{Fig2}(d). \\

\noindent {\bf Beyond quasiparticle pairs.---}  We now consider the tripartite mutual information $I_3$ between three adjacent intervals $A_1,A_2,A_3$ [cf.~Fig.~\ref{Fig1}(a)],
\begin{equation}
	I_{3}:=I_{2}[A_2,A_1]+I_{2}[A_2,A_3]-I_{2}[A_2,A_1\cup A_3]\, .
\label{Tripartite}
\end{equation}
This quantity allows us to discuss multipartite correlations --- $I_3$ is a fourpartite entanglement measure for pure states --- and to unveil peculiar features in the behavior of quantum correlations in the system.

By definition, the tripartite mutual information is zero if the mutual information between $A_2$ and $A_1\cup A_3$ 
is equal to the sum of the mutual information between $A_2$ and $A_1$ plus that between $A_2$ and $A_3$. This simple property allows us to argue that the mere presence of quasiparticle pairs 
must result in a vanishing tripartite mutual information. Indeed, pairs of 
quasiparticles can entangle only two subsystems at a time and different entangling pairs are uncorrelated with each other. This implies that 
the contributions of the pairs that entangle $A_2$ with $A_1$ and $A_3$ are subtracted by 
the last term in Eq.~\eqref{Tripartite}, so that $I_3=0$ in the scaling  
limit. In Fig.~\ref{Fig3}(a), as an example, we show how $I_3$ vanishes for the unitary dynamics implemented by $H$, when starting from the N\'eel state. 

Furthermore, in Ref.~\cite{SM} we demonstrate that not even triplets of quasiparticles can produce a finite tripartite mutual information, in the scaling limit. We also verified this numerically [see Fig.~\ref{Fig3}(b)] for the unitary dynamics implemented by $H$ starting from $\ket{\psi(0)}=\prod_{k} a^\dagger_{3k+1} \ket{0}$, which is a source of quasiparticle triplets \cite{bertini2018b}. On the other hand, multiplets with at least four elements can generate fourpartite entanglement [see sketch in Fig.~\ref{Fig1}(d)] giving rise to a nonvanishing $I_3/\ell$, in the $\ell\to\infty$ limit. This is shown in Fig.~\ref{Fig3}(c), obtained unitarily evolving the initial state $\ket{\psi(0)}=\prod_{k} a^\dagger_{4k+1} \ket{0}$, which is a source of quadruplets of quasiparticles \cite{bertini2018b}. 

We can thus exploit the tripartite mutual information to witness the existence of multiplets with at least four excitations in the process of Eq.~\eqref{eq-sto}. 
As shown in Fig.~\ref{Fig3}(d), the average tripartite mutual information $\overline{I}_3$ is indeed  different from zero. In particular, it assumes negative values, in contrast to the unitary dynamics shown in Fig.~\ref{Fig3}(c). We also observed that, for unitary dynamics, $I_3$ remains positive also in the presence of larger multiplets. This suggets that, in the Hamiltonian case, quantum information is shared in a redundant way among the different quasiparticles, i.e., they all share the same piece of information. On the other hand, in the presence of continuous monitoring we find $\overline{I}_3<0$, which indicates that the mutual information is monogamous \cite{hayden2013,rota2016,asadi2018,seshadri2018,ali-akbari2021}, $I_2[A_2,A_1 \cup A_3] >I_2[A_2,A_1]+I_2[A_2,A_3]$, and implies that there is more information about $A_2$ in $A_1 \cup A_3$ than in the sum of $A_1$ and $A_3$, separately. For the stochastic process, quantum information is thus highly delocalized, i.e., continuously and collectively distrubuted into the different excitations generated by the monitoring process and dispersed throughout the system by the coherent dynamics. \\

\noindent {\bf Conclusions.---} We have shown that the dynamics of quantum correlations in a paradigmatic continuously monitored system is unexpectedly complex and displays interesting unanticipated features. Such intricate phenomenology cannot be explained solely relying on entangled pairs of quasiparticles, and we have indeed provided evidence for the existence of multiplets of excitations with at least four elements.  From a fundamental perspective our results suggest that, if a quasiparticle picture for the process in Eq.~\eqref{eq-sto} exists, it cannot be solely based on an entangled-pair structure. We have further found that quantum correlations, including entanglement, can be enhanced by continuous monitoring [cf.~Fig.~\ref{Fig2}(c-d)]. The latter is also responsible for a robust delocalization of quantum information, which is shared by different subsystems in a genuinely collective way.

\noindent Finally, we note that quantum correlations in the average state, $\rho(t)=\mathbb{E}\left[\ket{\psi(t)}\!\bra{\psi(t)}\right]$, of noninteracting Lindblad dynamics are, in general, fully captured by a picture based on pairs of quasiparticles \cite{maity2020,alba2021,alba2021c,carollo2022}. \\

\noindent {\bf Acknowledgements.---} F.C.~acknowledges support from the “Wissenschaftler-R\"uckkehrprogramm GSO/CZS” of the Carl-Zeiss-Stiftung and the German Scholars Organization e.V., as well as through the Deutsche Forschungsgemeinsschaft (DFG, German Research Foundation) under Project No. 435696605, as well as through the Research Unit FOR 5413/1, Grant No. 465199066. F.C.~is indebted to the Baden-W\"urttemberg Stiftung for the financial support by the Eliteprogramme for Postdocs.

\bibliography{Notes_BIBLIO}

\setcounter{equation}{0}
\setcounter{figure}{0}
\setcounter{table}{0}
\makeatletter
\renewcommand{\theequation}{S\arabic{equation}}
\renewcommand{\thefigure}{S\arabic{figure}}

\makeatletter
\renewcommand{\theequation}{S\arabic{equation}}
\renewcommand{\thefigure}{S\arabic{figure}}

\onecolumngrid
\newpage

\setcounter{page}{1}

\begin{center}
{\Large SUPPLEMENTAL MATERIAL}
\end{center}
\begin{center}
\vspace{0.1cm}
{\Large Entangled multiplets and unusual spreading of quantum correlations in a continuously monitored tight-binding chain}
\end{center}
\begin{center}
Federico Carollo$^{1}$ and Vincenzo Alba$^2$
\end{center}
\begin{center}
$^1${\em Institut f\"ur Theoretische Physik, Universit\"at T\"ubingen,}\\
{\em Auf der Morgenstelle 14, 72076 T\"ubingen, Germany}\\
$^2$ {\em Dipartimento di Fisica, Universit\`a di Pisa, and INFN Sezione di Pisa, Largo Bruno Pontecorvo 3, Pisa, Italy}
\end{center}

\section{I. Prediction from the collapsed quasiparticle ansatz} 
From the collapsed quasiparticle ansatz ---which is a stochastic theory--- it is possible to obtain predictions for average quantities. Here, we show how to do this for entanglement-related quantities. In the first subsection, we discuss the case of a single subsystem. This result was derived in Ref.~\cite{cao2019} and we review it here also to clarify the scaling limit in which the theory should be expected  to be valid.  In the second subsection, we instead provide a new prediction, within the collapsed quasiparticle ansatz, for the mutual information and the logarithmic negativity between two subsystems embedded in an infinite chain. 

\subsection{A. The case of a single subsystem}
For a single subsystem of length $\ell$ and in the absence of continuous monitoring, the dynamics of the entanglement entropies is described by the following formula \cite{calabrese2005,alba2017,calabrese2018}
\begin{equation}
S_\ell^{(n),0}(t)\approx \int_{-\pi}^{\pi} \frac{dq}{2\pi} s_q^{(n)} \min \left\{ 2|v_q|t,\ell  \right\}\, .
\label{SM1}
\end{equation}
The above relation provides the leading-order behavior in $\ell$ of the entropy and is valid for large times. This can be made explicit by introducing the  scaling limit $\ell\to\infty$, with $t/\ell=\tau$ fixed. In this limit, we have  
\begin{equation}
{\bf {s}}^{(n),0}(\tau):=\lim_{\ell\to\infty} \frac{S_\ell^{(n),0}(\tau\ell)}{\ell}=\int_{-\pi}^{\pi} \frac{dq}{2\pi} s_q^{(n)} \min \left\{ 2|v_q|\tau,1  \right\}\, ,
\label{SM2}
\end{equation}
which is also the prediction used in the main text for the unitary quasiparticle picture. 

In Ref.~\cite{cao2019}, it was suggested that, in the presence of continuous monitoring, the entanglement entropies, averaged over all possible realizations of the stochastic process, obey the following relation 
\begin{equation}
\overline{S}^{(n)}_\ell(t)=\mathbb{E}\left[S_\ell^{(n)}(t)\right]\approx e^{-\gamma t } \int_{-\pi}^{\pi}  \frac{dq}{2\pi}  s_q^{(n)}\min \left\{ 2|v_q|t,\ell  \right\}+\gamma \int_0^t du \,  e^{-\gamma u}     \int_{-\pi}^{\pi} \frac{dq}{2\pi} s_q^{(n)} \min \left\{ 2|v_q|u,\ell  \right\}\, .
\label{SM3}
\end{equation}
The above formula is specialized for the case of a (macroscopically) homogeneous initial state, such as the one considered in the main text. A straightforward calculation leads to the following simplified expression 
\begin{equation}
\overline{S}^{(n)}_\ell(t)\approx \int_{-\pi}^{\pi}  \frac{dq}{2\pi}  \frac{2|v_q|}{\gamma}\left(1-e^{-\gamma \min\left\{t,\frac{\ell}{2|v_q|}\right\}}\right)s_q^{(n)} \, .
\label{SM4}
\end{equation}
It is important to look at this formula in the scaling limit. This is indeed essential to systematically compare the prediction from the theory with numerical simulations and to establish the validity of the theory. 

The definition of a proper limit for the above formula requires a bit of care. Indeed, one should note that, in the scaling limit in which the quasiparticle picture becomes well-defined, time diverges as $t=\tau\ell\to\infty$. As such, the exponential function would converge to $0$ as soon as $\tau>0$, unless a proper rescaling of $\gamma$ is introduced. This is also shown by numerical results presented in Fig.~\ref{Fig_SM}(a). Clearly, by looking at the arguments of the $\min$ function, the correct scaling of $\gamma$ is $\gamma=\Gamma/\ell$, with $\Gamma$ independent on $\ell$. Finally, due to the appearence of a factor $\gamma^{-1}$ in the integral, a well-defined limit is obtained by dividing $\overline{S}^{(n)}_\ell(t)$ by $\ell$. Note that this scaling limit is the most natural one, emerging when considering hydrodynamic theories in combination with dissipative effects. It was also recognized in Ref.~\cite{cao2019} but it is actually the same, in spirit, dissipative limit considered, e.g., in Refs.~\cite{lange2018,bouchoule2020}, where time is take in units of the (small) dissipative rate, i.e.~$\gamma t$ is kept fixed. 

Collecting these observations, we define the scaling limit $\ell\to\infty$, with $t/\ell=\tau$ and $\gamma t=\Gamma \tau$ fixed. In this limit, we find 
\begin{equation}
{\bf {s}}^{(n)}(\tau)=\lim_{\ell\to\infty}\frac{\overline{S}_\ell^{(n)}(t)}{\ell}= \int_{-\pi}^{\pi}  \frac{dq}{2\pi}  \frac{2|v_q|}{\Gamma}\left(1-e^{-\Gamma \min\left\{\tau,\frac{1}{2|v_q|}\right\}}\right)s_q^{(n)} \, ,
\label{SM5}
\end{equation}
which is the scaling function predicted by the collapsed quasiparticle ansatz and plotted in Fig.~\ref{Fig2}(a-b). We note that, for our initial state we have $s^{(n)}_q =\log 2$ for all $n$, and the same quantitative prediction holds for all entanglement entropies.

\subsection{B. The case of two (adjacent) subsystems}
We now focus on the prediction, derived from the assumptions of the collapsed quasiparticle ansatz, for quantum correlations and entanglement between two adjacent subsystems of length $\ell$.  

We start considering the mutual information between these subsystems. As for the previous case, the starting point is the prediction for the unitary case, given by 
\begin{equation}
I_2^{0}(t)\approx \int_{-\pi}^{\pi}  \frac{dq}{2\pi}  \Theta_q(t) \log 2 \, ,
\label{SM6}
\end{equation}
where the factor $\log 2$ comes from considering the correlation content of quasiparticles which is given by the von Neumann Yang-Yang entropy. The function $\Theta_q(t)$ is here given by 
\begin{equation}
\Theta_q(t)=2|v_q|t+2\max\left\{|v_q|t,\ell\right\}-2\max\left\{2|v_q|t,\ell\right\}\, .
\label{SM7}
\end{equation}
With the above formula, we can directly write down the prediction from the collapsed quasiparticle ansatz. In the same spirit of Eq.~\eqref{SM3}, we have  
\begin{equation}
\overline{I}_2(t)=\mathbb{E}\left[I_2(t)\right]\approx e^{-\gamma t}\int_{-\pi}^{\pi}  \frac{dq}{2\pi}  \Theta_q(t)\log 2 +\gamma \int_0^{t} du \, e^{-\gamma u} \int_{-\pi}^{\pi}  \frac{dq}{2\pi}  \Theta_q(u)\log 2 \, .
\label{SM8}
\end{equation}
A lengthy but straightforward calculation gives
\begin{equation}
\overline{I}_2(t)\approx \log 2\int_{-\pi}^{\pi}\frac{dq}{2\pi} \left[e^{-\gamma t} \Theta_q(t)+J_1\left(\min\left\{t,\frac{\ell}{2|v_q|}\right\}\right)+J_2\left(\min\left\{t,\frac{\ell}{|v_q|}\right\},\min\left\{t,\frac{\ell}{2|v_q|}\right\}\right)\right]\, ,
\label{SM9}
\end{equation}
where we have 
$$
J_1(t)=\gamma \int_0^t du \, e^{-\gamma u} \, 2|v_q| u\,  \qquad \mbox{ and } \qquad J_2(t,s)=\gamma \int_s^t du\, e^{-\gamma u}\left[2\ell -2|v_q| u\right]\, .
$$
When properly considering the scaling limit discussed in the previous section, the prediction becomes 
\begin{equation}
\lim_{\ell\to\infty}\frac{\overline{I}_2(\tau \ell)}{\ell}=\log 2 \int_{-\pi}^{\pi}\frac{dq}{2\pi} \left[e^{-\Gamma \tau} \Theta_q(t)+j_1\left(\min\left\{\tau,\frac{1}{2|v_q|}\right\}\right)+j_2\left(\min\left\{\tau,\frac{1}{|v_q|}\right\},\min\left\{\tau,\frac{1}{2|v_q|}\right\}\right)\right]\, ,
\label{SM10}
\end{equation}
with the rescaled integrals
$$
j_1(\tau)=\Gamma \int_0^\tau du \, e^{-\Gamma u} \, 2|v_q| u\, , \qquad \mbox{ and } \qquad j_2(\tau,\sigma)=\Gamma \int_\sigma^\tau du\, e^{-\Gamma u}\left[2 -2|v_q| u\right]\, .
$$
The formula in Eq.~\eqref{SM10} is the one plotted in Fig.~\ref{Fig2}(c) for the collapsed quasiparticle ansatz. Furthermore, from the above formula we can also find the stationary value for the mutual information. This can be found by taking the limit $\tau\to\infty$, which gives 
\begin{equation}
	\lim_{\tau\to\infty}\lim_{\ell\to\infty} \frac{\overline{I}_2(\tau\ell)}{\ell}=\frac{\log 2}{\Gamma}\int_{-\pi}^\pi\frac{dq}{\pi}\left(1-e^{-\Gamma/(2|v_q|)}\right)^2|v_q|.
\end{equation}

We now briefly comment on the prediction for the logarithmic negativity between the two adjacent intervals. We first observe that the logarithmic negativity for the unitary case, $\mathcal{E}^{0}_2$, has been shown to be equivalent to half of the mutual information, i.e., $\mathcal{E}^{0}_2=I_2^{0}/2$, since in our case the latter is equal to the R\'enyi-$1/2$ mutual information. Because of this, the prediction for the logarithmic negativity in the continuously monitored system is immediately found as $\overline{\mathcal{E}}_2=\overline{I}_2/2$. 

\section{II. Details on the numerical simulations } 
The numerical simulations of quantum trajectories were performed using the method discussed in Refs.~\cite{cao2019,alberton2021}. The curves plotted in Fig.~\ref{Fig2} and in Fig.~\ref{Fig3}(d) have been obtained by simulating a system with $L=1000$ sites and $\ell$ up to $\ell=100$. We have discretized time using a time step $J dt= 0.025 $ and we have verified that our results are not substantially changed by considering a larger $Jdt= 0.05 $ nor a smaller one, $J dt= 0.01 $ [see e.g.~Fig.~\eqref{Fig_SM}(c)]. The curves are obtained by taking the average over a different number of quantum trajectories of the stochastic process [see caption of Fig.~\ref{Fig2} and Fig.~\ref{Fig3}]. The plots in Fig.~\ref{Fig3}(a-b-c) are for $\gamma\equiv 0$. These curves do not require averaging over trajectories since the dynamics is deterministic and are also free of time-discretization (Trotter) errors.

\section{III. Additional results} 

\begin{figure}[t]
\centering
\includegraphics[width=\textwidth]{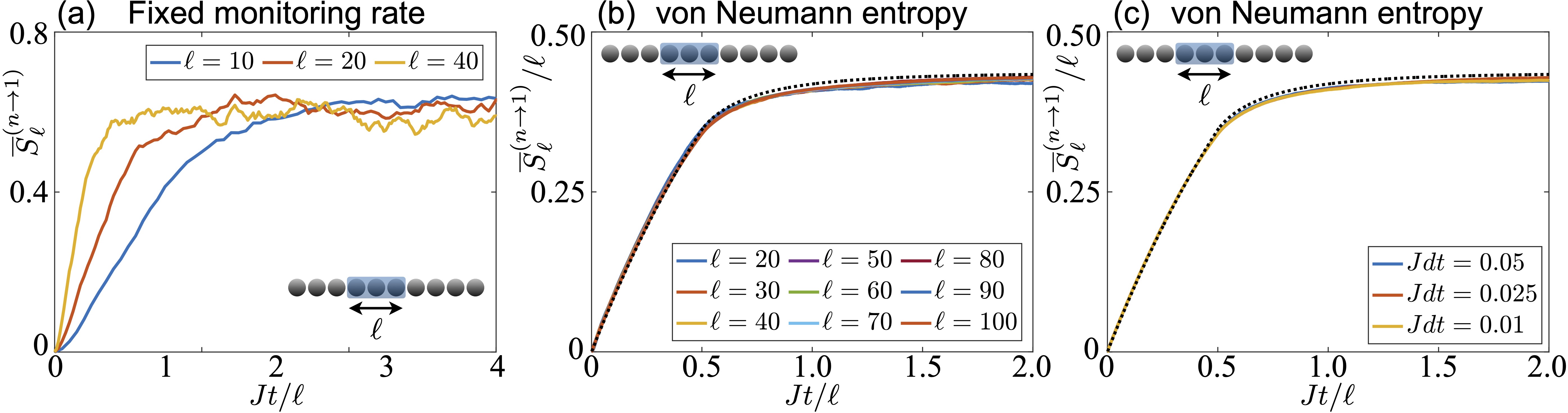}
\caption{ {\bf Additional results.} (a) Plot of the von Neumann entanglement entropy for a subsystem of length $\ell$ embedded in the many-body chain. The plot is done in the scaling limit in which the unitary quasiparticle picture is expected to emerge. Note however that we are not dividing the entropy by $\ell$. In this case, we consider a fixed $\gamma=J$.  The curves we display, for $\ell=10,20,40$, show that in this case there is no scaling behavior. The entropy is averaged over $N_{\rm traj}=200$ trajectories. (b) Comparison of numerical results for the von Neumann entanglement entropy (in the scaling limit $t,\ell\to\infty$ with $t/\ell=\tau$ fixed and $\gamma \ell=\Gamma$ fixed) and the prediction from the collapsed quasiparticle ansatz (dotted line). Here, we set $\Gamma=1$. While the prediction is closer to the numerical results than what we observed for the other entropies shown in the main text, it seems that a perfect agreement is not reached. The entropy is averaged over $N_{\rm traj}=250$ and here we considered $Jdt=0.025$. (c) Plot of the von Neumann entanglement entropy in the scaling regime, for different values of $J dt$. As shown, for the considered values of $J dt$, it is not possible to appreciate relevant changes in the behavior of the von Neumann entropy. We consider $N_{\rm traj}=50$ for $Jdt=0.05$, $N_{\rm traj}=200$ for $Jdt=0.025$ and $N_{\rm traj}=100$ for $Jdt=0.01$. 
}
\label{Fig_SM}
\end{figure}

In this section, we make two important considerations. First, we show that if $\gamma$ is not rescaled as $\gamma=\Gamma/\ell$ then the entanglement entropies do not show scaling behavior when considering the limit $t,\ell\to\infty$ with $t/\ell$ fixed. This can be observed in Fig.~\ref{Fig_SM}(a), where we plot the entanglement entropy $S^{(n\to 1)}_\ell$ as a function of $J t/\ell$. As shown, increasing $\ell$ the entanglement entropy tends to converge faster and faster to its stationary behavior and, thus, it cannot show a proper scaling behavior. This, in addition to the discussion in the main text, shows that in order to arrive at a proper scaling limit, $\gamma$ needs to be rescaled in a way that $\gamma t$ is also fixed in the limit. 

We further show numerical results, in the appropriate scaling limit discussed in the main text and in this supplemental material, for the von Neumann entanglement entropy investigated in Ref.~\cite{cao2019}. We can observe that, also in this case, the numerical results seem not to converge to the prediction from the collapsed quasiparticle ansatz, even if, in this case, the prediction is much more accurate. This is shown in Fig.~\ref{Fig_SM}(b). In Fig.~\eqref{Fig_SM}(c), we show that the discrepancy found in Fig.~\eqref{Fig_SM}(b), between numerical results and collapsed quasiparticle ansatz, does not seem to be related to a time-discretization Trotter error.

\section{IV. Vanishing of tripartite information with triplets} 
In this section we consider the state  $\prod_{j=0}^{L/\nu-1}a^\dagger_{\nu j+1}|0\rangle=|10\dots0100100\dots\rangle$. For $\nu=3$, this state is a source of triplets of quasiparticles for the considered Hamiltonian and here we show that, in this case,
the tripartite information $I_3$ vanishes in the scaling limit. 

For the above state, $\nu$ species of quasiparticles exist. They have group 
velocities $v_p^j$ 
\begin{equation}
	v_p^j=J\sin\left(p-\frac{2(j-1)\pi}{\nu}\right), \quad\mathrm{with}\,\,j=1,2,\dots \nu\, .
\end{equation}
Here $j$ labels the different species of quasiparticles. The quasimomentum $k^j_p$ of the 
quasiparticles is 
\begin{equation}
	k^j_p=p-\frac{2(j-1)\pi}{\nu}\, , 
\end{equation}
where 
\begin{equation}
	p\in \left[\pi-\frac{2\pi}{\nu},\pi\right]. 
\end{equation}
For $\nu=2$ one recovers the standard case of entangled pairs of quasiparticles. 
Now the two species of quasiparticles have $p\in[0,\pi]$ and quasimomenta 
$k^1_p\in[0,\pi]$, whereas $k^2_p\in[-\pi,0]$. This means that $v_p^1=-v_p^2$. 

\begin{figure}[b]
\centering
\includegraphics[width=0.7\textwidth]{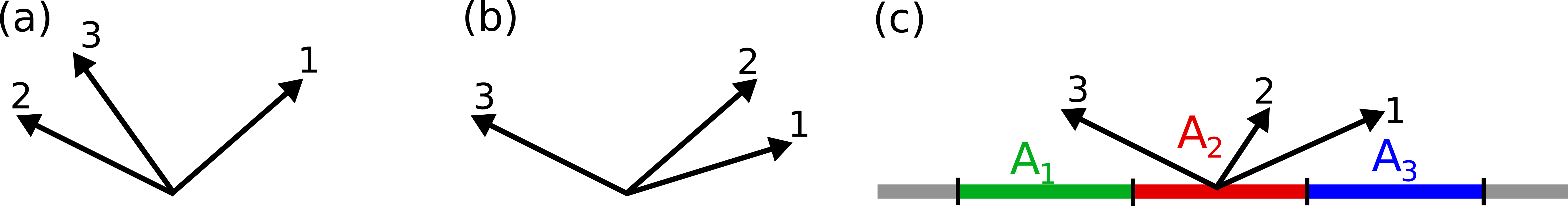}
\caption{ {\bf Entangling triplets.} We show two possible quasiparticles 
 configurations. In (a) quasiparticles $2,3$ have negative velocities, 
 whereas quasiparticle $1$ has  positive one. This corresponds to 
 $\pi/3\le p\le \pi/2$. In (b) quasiparticles 
 $1,2$ have positive velocities, and $3$ has a negative one. This 
 corresponds to $2/3\pi\le p\le 5/6\pi$. Two other 
 equivalent configurations are obtained by exchanging $2\leftrightarrow 3$ 
 in (a) and $1\leftrightarrow 2$ in (b), which give $\pi/2\le p\le 2/3\pi$ and 
 $5/6\pi\le p\le \pi$, respectively. (c) Configuration with quasiparticles of a same triplet shared by three subsystems, $A_1, A_2$ and $A_3$. 
}
\label{fig:triplets}
\end{figure}

Now for the case with $\nu=3$ one has different possible orderings for the 
quasiparticles velocities as 
\begin{align}
	 v^2_p<v^3_p<0, v^1_p>0 & &\frac{\pi}{3}\le p<\frac{\pi}{2}\\
	 v^3_p<v^2_p<0, v^1_p>0 & &\frac{\pi}{2}\le p<\frac{2\pi}{3}\\
	 v^3_p<0, v^2_p>0,v_p^2<v^1_p & &\frac{2\pi}{3}\le p<\frac{5\pi}{6}\\
	 v^3_p<0, v^1_p>0,v^2_p>v^1_p & &\frac{5\pi}{6}\le p<\pi
 \end{align}
This means that the triplets of quasiparticles for a generic label $p$ can 
be arranged as in Fig.~\ref{fig:triplets}. 
%
%

Let us now discuss the contribution of the triplets to the entanglement entropy. 
Crucially, this depends in the specific way in which the entangled quasiparticles 
are shared. Let us consider two subsystems $X$ and $Y$. The contribution of a 
$\nu$-plet of entangled quasiparticles to the 
entanglement entropy between them is denoted as $s^{\{j_i\}}_p$. 
Here $j_i=1,2,\dots,\nu$ and $i=0,1,\dots,m$ with $m$ being the number of 
quasiparticles in $X$. It has been postulated in Ref.~\cite{bertini2018b} that 
\begin{equation}
	s^{\{j_i\}}_p=-\left(\sum_{i=1}^m\varrho^{(j_i)}_p\right)
	\ln\left(\sum_{i=1}^m\varrho^{(j_i)}_p\right)
-\left(1-\sum_{i=1}^m\varrho^{(j_i)}_p\right)
	\ln\left(1-\sum_{i=1}^m\varrho^{(j_i)}_p\right). 
\end{equation}
Here the densities $\varrho^{(j_i)}_p$ are defined as 
\begin{equation}
	\varrho^{(j)}_p=\varrho\left(p-\frac{2(j-1)\pi}{\nu}\right), 
\end{equation}
where $\varrho(k)$ is the density describing the stationary value of 
local observables, and it is calculated as 
\begin{equation}
\varrho(k)=\langle\psi(0)|\beta^\dagger_k \beta_k|\psi(0)\rangle. 
\end{equation}
Importantly, one has that the entropy contribution satisfies 
\begin{equation}
	s^{\{j_i\}}_p=s^{\{j_i\}_c}_p,\quad \mathrm{with}\,\,\{j_i\}_c=\{1,2,\dots,\nu\}/\{j_i\}. 
\end{equation}
This means that $s^{\{j_i\}}_p$ is symmetric under exchange of the particles 
that are in and out of a subsystem. This property ensures that the entanglement entropy
of two complementary subsystems is the same, i.e., $S_X=S_{Y}$ if $Y=\overline{X}$. 
As we show, this also implies a vanishing tripartite information if only entangled triplets are 
present. 

For the case of triplets we have the contributions 
\begin{align}
	\label{eq:1}
	& s^{\{1\}}_p=s^{\{2,3\}}_p\\
	\label{eq:2}
	& s^{\{2\}}_p=s^{\{1,3\}}_p\\
	\label{eq:3}
	& s^{\{3\}}_p=s^{\{1,2\}}_p. 
\end{align}
The only arrangement of quasiparticles for which one might expect a non-zero tripartite mutual information is that with the triplet shared by the three subsystems. 
This is a configuration like the one shown in Fig.~\ref{fig:triplets} (c). System $A$ is 
tripartite as $A=A_1\cup A_2\cup A_3$. In the configuration we have 
that $A_1$ contains a quasiparticle of species $3$, $A_2$ one of species 
$2$, and $A_3$ of species $1$. Notice that although there could be 
entangled multiplets shared between the subsystems and the remainder 
of $A$, we can neglect them because they cancel when constructing the 
mutual information. Let us consider the tripartite mutual information 
\begin{equation}
	\label{eq:tri}
	I_3=I_2[A_2,A_1]+I_2[A_2,A_3]-I_2[A_2,A_1\cup A_3]. 
\end{equation}
Now it is clear that 
\begin{align}
	\label{eq:s1}
	&I_2[A_2,A_1]=s^{\{2\}}_p+s^{\{3\}}_p-s^{\{2,3\}}_p\\
	\label{eq:s2}
	&I_2[A_2,A_3]=s^{\{2\}}_p+s^{\{1\}}_p-s^{\{1,2\}}_p\\
	\label{eq:s3}
	&I_2[A_2,A_1\cup A_3]=s^{\{2\}}_p+s^{\{1,3\}}_p,
\end{align}
where in the last row we used that $s^{\{1,2,3\}}=0$. 
Now, after substituting Eqs.~\eqref{eq:s1}-\eqref{eq:s2}-\eqref{eq:s3} in Eq.~\eqref{eq:tri}, 
by using Eqs.~\eqref{eq:1}-\eqref{eq:2}-\eqref{eq:3}, we find that 
$I_3$ vanishes. This argument can be adapted to any triplet 
arrangement.

\end{document}